\documentclass[english,aps, prl, twocolumn, superscriptaddress, showpacs]{revtex4}
\pdfoutput=1
\usepackage[latin9]{inputenc}
\usepackage{amsmath}
\usepackage{graphicx}
\usepackage{amssymb}
\usepackage{esint}

\usepackage{babel}

\begin{document}

\title{The photon shuttle: Landau-Zener-Stueckelberg dynamics\\
 in an optomechanical system}

\author{Georg Heinrich}

\affiliation{Arnold Sommerfeld Center for Theoretical Physics, Center for NanoScience
and Department of Physics, Ludwig-Maximilians-Universität München,
Theresienstr. 37, D-80333 München, Germany}

\author{J. G. E. Harris}

\affiliation{Department of Physics, Yale University, 217 Prospect Street, New
Haven, Connecticut, 06520, USA}

\affiliation{Department of Applied Physics, Yale University, 15 Prospect Street,
New Haven, Connecticut 06520, USA}

\author{Florian Marquardt}

\affiliation{Arnold Sommerfeld Center for Theoretical Physics, Center for NanoScience
and Department of Physics, Ludwig-Maximilians-Universität München,
Theresienstr. 37, D-80333 München, Germany}
\begin{abstract}
The motion of micro- and nanomechanical resonators can be coupled
to electromagnetic fields. Such optomechanical setups allow one to
explore the interaction of light and matter in a new regime at the
boundary between quantum and classical physics. We propose an approach
to investigate non-equilibrium photon dynamics driven by mechanical
motion in a recently developed setup with a membrane between two mirrors,
where photons can be shuttled between the two halves of the cavity.
For modest driving strength we predict the possibility to observe
an Autler-Townes splitting indicative of Rabi dynamics. For large
drive, we show that this system displays Landau-Zener-Stueckelberg
dynamics originally known from atomic two-state systems. 
\end{abstract}


\maketitle
Landau-Zener (LZ) transitions \cite{Landau1932On-the-theory-o,Zener1932Non-Adiabatic-C}
are essential to the dynamics of many physical systems. In the usual
model, a parameter in a two-state Hamiltonian is swept through an
avoided level crossing where the two bare eigenstates $|1\rangle$
and $|2\rangle$ hybridize. When the parameter is changed at a finite
speed, the system may undergo a LZ transition into the other eigenstate.
Beyond this standard LZ problem, the dynamics becomes more elaborate
if repeated transitions are take into account. For a periodic modulation
of the parameter, the first LZ transition splits the state into a
coherent superposition $\alpha|1\rangle+\beta|2\rangle$. Due to the
difference in energy, the system afterwards accumulates a relative
phase between states $|1\rangle$ and $|2\rangle$. Thus, when returned
to the avoided crossing, the system undergoes quantum interference
with itself during the second LZ transition. This leads to interference
patterns for the state population, so called Stueckelberg oscillations
\cite{Stuckelberg1932Theorie-der-une}. Originally, Landau-Zener-Stueckelberg
(LZS) dynamics was studied in atomic systems \cite{Baruch1992Ramsey-interfer,Yoakum1992Stueckelberg-os,Mark2007Stuckelberg-Int}.
Recently, the concept has been applied to superconducting qubits \cite{Oliver12092005}.
Currently, there is growing interest in LZ and LZS dynamics concerning
topics such as state preparation and entanglement \cite{Saito2006Quantum-state-p,Wubs2006Gauging-a-Quant},
cooling or qubit spectroscopy \cite{Berns2008Amplitude-spect}. Another
rapidly evolving area of research is optomechanics (see \cite{Marquardt2009Optomechanics}
for a recent review and further references). Optomechanical systems
couple mechanical degrees of freedom to radiation fields. This provides
new means to manipulate both the light field and the mechanical motion.
Apart from the hope to eventually explore the quantum regime of mechanical
motion, there have been several studies of the complex nonlinear dynamics
of these systems \cite{Metzger2008Self-Induced-Os,Marquardt2006Dynamical-Multi,Ludwig2008The-optomechani,Kippenberg2005Analysis-of-Rad}.

Here, we propose an approach to observe dynamics of the light field
in a driven optomechanical system, in the form of LZS oscillations.
We note that there exist some purely optical setups \cite{Spreeuw1990Classical-reali,Bouwmeester1995Observation-of-}
that have mimicked quantum two-state and standard LZ dynamics (but
not LZS oscillations). In the optomechanical setup analyzed here,
the mechanical motion of a membrane placed between two fixed mirrors
is driven such that the resulting motion shuttles photons between
the two halves of the cavity. A setup of this kind was recently realized
in \cite{ThompsonStrong-dispersi,Jayich2008Dispersive-opto}. %
\begin{figure}
\includegraphics[width=1\columnwidth]{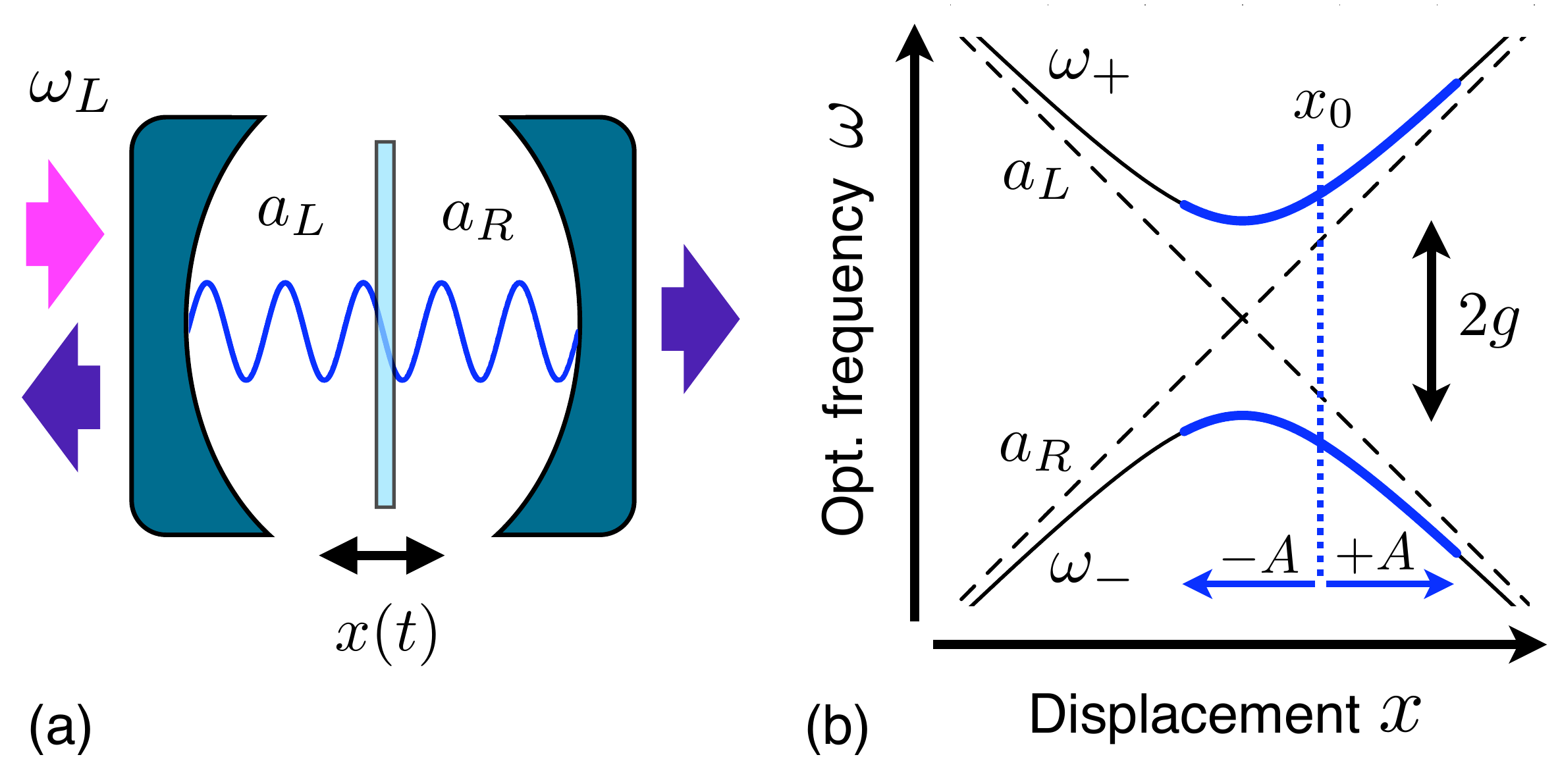}

\caption{(a) Setup: a dielectric membrane couples two modes $a_{L}$, $a_{R}$
inside a cavity. The left hand side is excited by a laser $\omega_{L}$
while the transmission to the right is recorded. (b) Optical resonance
frequency as function of displacement: the membrane's displacement
linearly changes the bare mode frequencies (dashed). Due to the coupling
$g$, there is an avoided crossing of the eigenfrequencies $\omega_{\pm}$
(black). The membrane is driven, with $x(t)=A\cos(\Omega t)+x_{0}$
(blue).\label{fig:Setup}}

\end{figure}

We consider two cavity modes coupled by a dielectric membrane placed
in the middle between two high-finesse mirrors, see Fig. \ref{fig:Setup}a.
The system Hamiltonian reads

\begin{eqnarray}
\hat{H}_{{\rm sys}} & = & \hbar\omega_{0}\left(1-\frac{x(t)}{l}\right)\hat{a}_{L}^{\dagger}\hat{a}_{L}+\hbar\omega_{0}\left(1+\frac{x(t)}{l}\right)\hat{a}_{R}^{\dagger}\hat{a}_{R}\nonumber \\
 & + & \hbar g\left(\hat{a}_{L}^{\dagger}\hat{a}_{R}+\hat{a}_{R}^{\dagger}\hat{a}_{L}\right)+\hat{H}_{{\rm drive}}+\hat{H}_{{\rm decay}}.\label{eq:H_sys}\end{eqnarray}
$\hat{a}_{L}^{\dagger}\hat{a}_{L}$ and $\hat{a}_{R}^{\dagger}\hat{a}_{R}$
are the photon numbers for the two optical modes in the left and right
cavity half (each of length $l$), respectively, whose resonance frequency
$\omega_{0}$ is changed due to the displacement $x$ of the membrane.
The coupling $g$ describes the photon tunneling through the membrane.
Due to the coupling, there is an avoided crossing in the optical resonance
frequency $\omega_{\pm}(x)=\pm\sqrt{g^{2}+(\omega_{0}x/l)^{2}}$,
see Fig.~\ref{fig:Setup}b. We propose to drive the membrane with
mechanical frequency $\Omega$ and resulting amplitude $A$ around
a mean position $x_{0}$,\begin{equation}
x(t)=A\cos(\Omega t)+x_{0},\label{eq:drive_displacement}\end{equation}
and investigate the system in the regime where the timescale of photon
exchange is comparable to the timescale of the mechanical motion ($g\simeq\Omega$).
Recently the coupling frequency $g/2\pi$ has been significantly reduced
by exploiting properties of transverse modes \cite{Sankey2009Improved-Positi},
and it is tunable down to $200\,\text{kHz}$ at present. The mechanical
eigenfrequencies of typical $1\,\text{mm\ensuremath{\times}1\ensuremath{\,}\text{mm\ensuremath{\times}50\ensuremath{\,}\text{nm}}}$
membranes range between $100\,\text{kHz}$ and $1\,\mbox{MHz}$. Commercially
available membrane sizes should allow this to go from $20\,\text{kHz}$
up to $10\,\text{MHz}$. We point out that here $\Omega$ need not
coincide with the membrane's eigenfrequency but depends only on the
driving.

We assume the left hand side of the cavity is driven by a laser of
frequency $\omega_{L}$ and amplitude $b^{in}$.Our goal is to examine
the photon dynamics by looking at the transmission $T$. Using input/output
theory, the equations of motion for the average fields $a_{L}=\left\langle \hat{a}_{L}\right\rangle $
and $a_{R}=\left\langle \hat{a}_{R}\right\rangle $ read\begin{eqnarray}
\frac{d}{dt}a_{L} & = & \frac{1}{i}\left[-\bar{x}(t)\, a_{L}+g\, a_{R}\right]-\frac{\kappa}{2}a_{L}-\sqrt{\kappa}\, b_{L}^{in}(t)\nonumber \\
\frac{d}{dt}a_{R} & = & \frac{1}{i}\left[+\bar{x}(t)\, a_{R}+g\, a_{L}\right]-\frac{\kappa}{2}a_{R},\label{eq:EOM_IO_theory}\end{eqnarray}
with the cavity decay rate $\kappa$ for each of the modes, and the
drive $b_{L}^{in}(t)=e^{-i\Delta_{L}t}b^{in}$. Here, we used a rotating
frame, with laser detuning from resonance $\Delta_{L}=\omega_{L}-\omega_{0}$.
The displacement is written in terms of a frequency via $\bar{x}(t)=(\omega_{0}/l)x(t)$,
likewise for $\bar{A}$, $\bar{x}_{0}$. The transmission to the right,
$T(t)=\kappa\langle\hat{a}_{R}^{\dagger}(t)\hat{a}_{R}(t)\rangle/(b^{in})^{2}$,
can be expressed as\begin{equation}
T(t)=\kappa^{2}\left|\int_{-\infty}^{t}G(t,t')\, e^{-i\Delta_{L}t'-(\kappa/2)(t-t')}\, dt'\right|^{2},\label{eq:T_Greens_fct}\end{equation}
where the phase factor includes laser drive and cavity decay, while
the Green's function $G(t,t')$ describes the amplitude for a photon
to enter the cavity from the left at time $t'$ and to be found in
the right cavity mode later at time $t$. Technically, $G(t,t')$
is found by setting $\kappa=0$ in Eq.~(\ref{eq:EOM_IO_theory})
and solving for $a_{R}(t)$ with the initial conditions $a_{L}(t')=1$,
$a_{R}(t')=0$. 

\begin{figure}
\includegraphics[width=1\columnwidth]{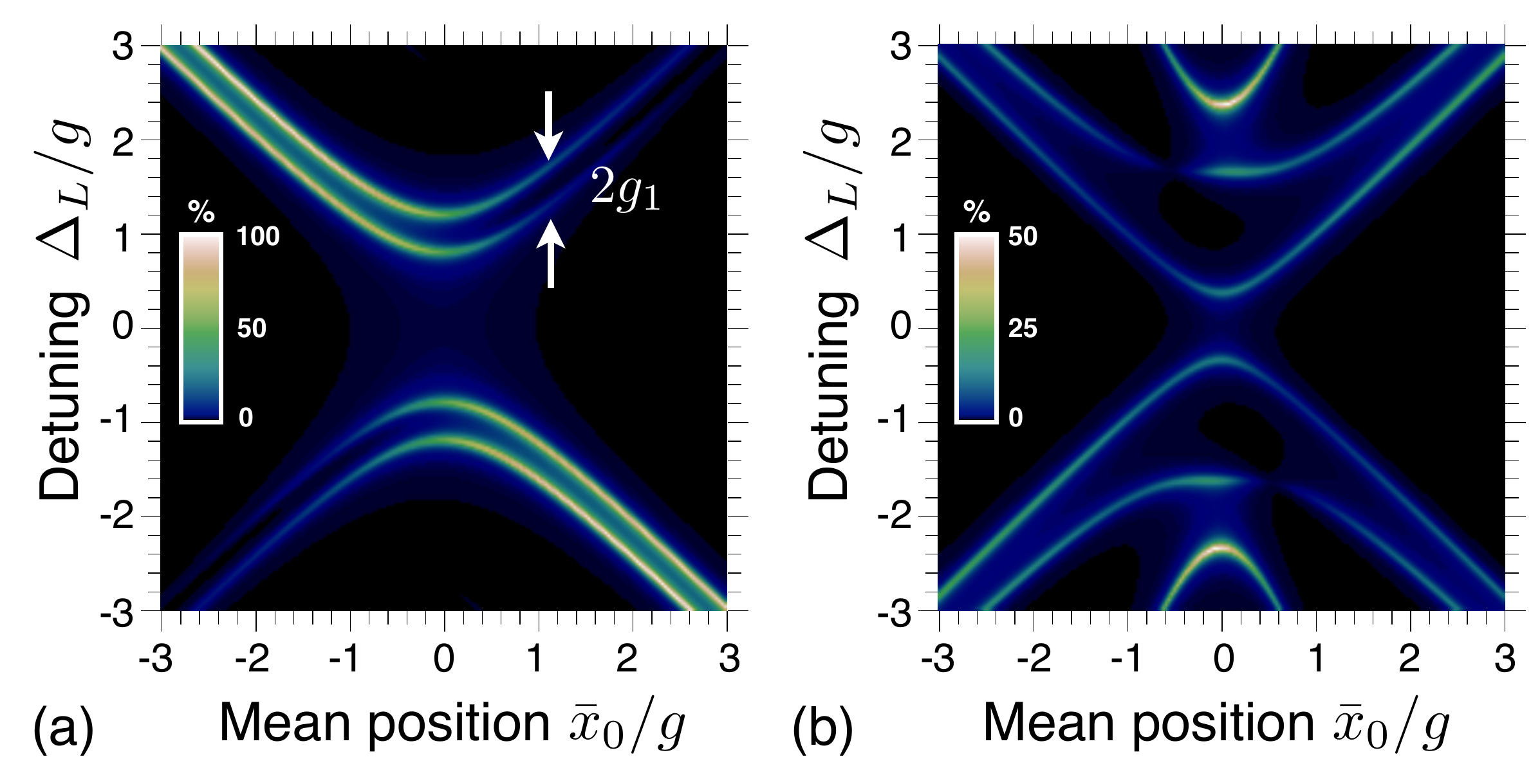}

\caption{(a) Autler-Townes splitting of the cavity frequencies $\omega_{\pm}$
due to the mechanical motion: density plot for the time-averaged transmission
depending on $\bar{x}_{0}$ and $\Delta_{L}$. For every position
$\bar{x}_{0}$ the mechanical drive frequency is set to be $\Omega=2\sqrt{g^{2}+\bar{x}_{0}^{2}}$
such that it is always resonant with the difference between the two
optical mode frequencies $\omega_{\pm}$. Further parameters are amplitude
$\bar{A}=0.2\,\Omega$ and decay $\kappa=0.1\, g$. The splitting
is set by the Rabi frequency $g_{1}\simeq g\bar{A}/\Omega$, proportional
to the drive amplitude. (b) Transmission as in (a) but for stronger
drive $\bar{A}=1.6\,\Omega$. Mechanical sidebands, displaced by $\pm\Omega$,
become visible and interact. \label{fig:AutlerTownes}}

\end{figure}

We start investigating the dynamics by considering modest drive amplitudes
$\bar{A}<\Omega$. Fig.~\ref{fig:AutlerTownes}a displays the time-averaged
transmission depending on $\bar{x}_{0}$ and $\Delta_{L}$. We observe
an Autler-Townes splitting \cite{Autler1955Stark-Effect-in,Spreeuw1990Classical-reali}
of the two hyperbola branches $\omega_{\pm}$. Indeed, the mechanical
drive induces Rabi oscillations between the two photon branches, at
a Rabi frequency $g_{1}\simeq g\bar{A}/\Omega$, leading to a corresponding
splitting in the spectroscopic picture. For larger drive amplitudes
the dynamics becomes more involved. For instance mechanical sidebands
arise as shown in Fig. \ref{fig:AutlerTownes}b, and they start to
interact with each other. In the following, we will focus on the dynamics
of this strong driving regime.%

\begin{figure}
\begin{centering}
\includegraphics[width=1\columnwidth]{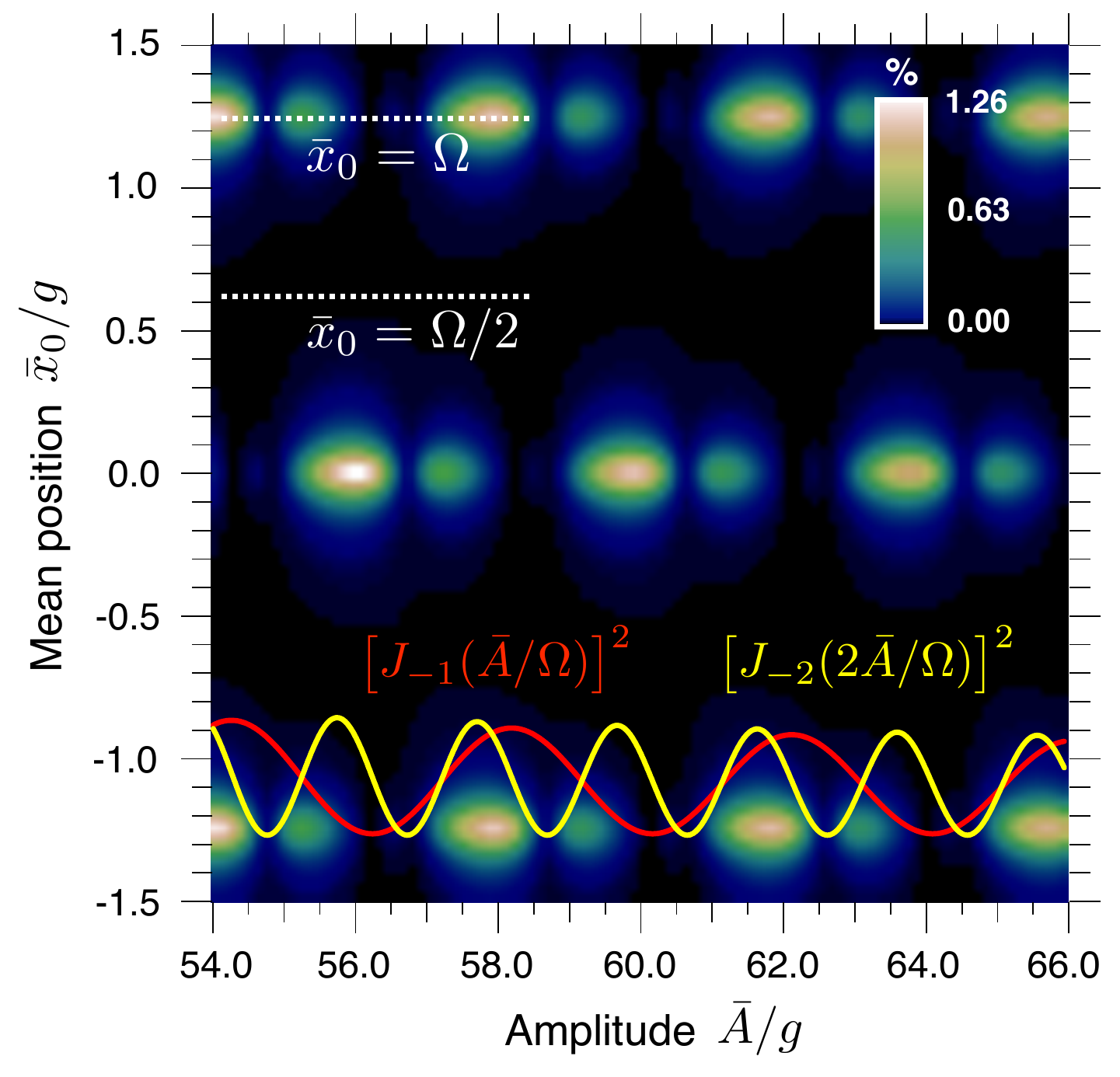}
\par\end{centering}

\caption{\label{fig:Large_amplitudes}Density plot for the time-averaged transmission
as function of average displacement $\bar{x}_{0}$ and mechanical
drive amplitude $\bar{A}\gg\Omega,\, g$. Further parameters are laser
detuning $\Delta_{L}=0$, mechanical frequency $\Omega/2\pi=0.2g$
and cavity decay $\kappa=0.2g$. Finite transmission is observed when
the resonance conditions Eq. (\ref{eq:Resonance_cond1}) and (\ref{eq:Resonance_cond2})
for multiphonon transitions are met. The transmission is modulated
according to the product of two Bessel functions. For the case $\bar{x}_{0}=-\Omega$
both are depicted in the plot's plane. Red: $\sim J_{m}^{2}(\bar{A}/\Omega)$,
due to the excitation process. Yellow: $\sim J_{n}^{2}(2\bar{A}/\Omega)$,
due to LZS dynamics.}

\end{figure}

\begin{figure}
\begin{centering}
\includegraphics[width=1\columnwidth]{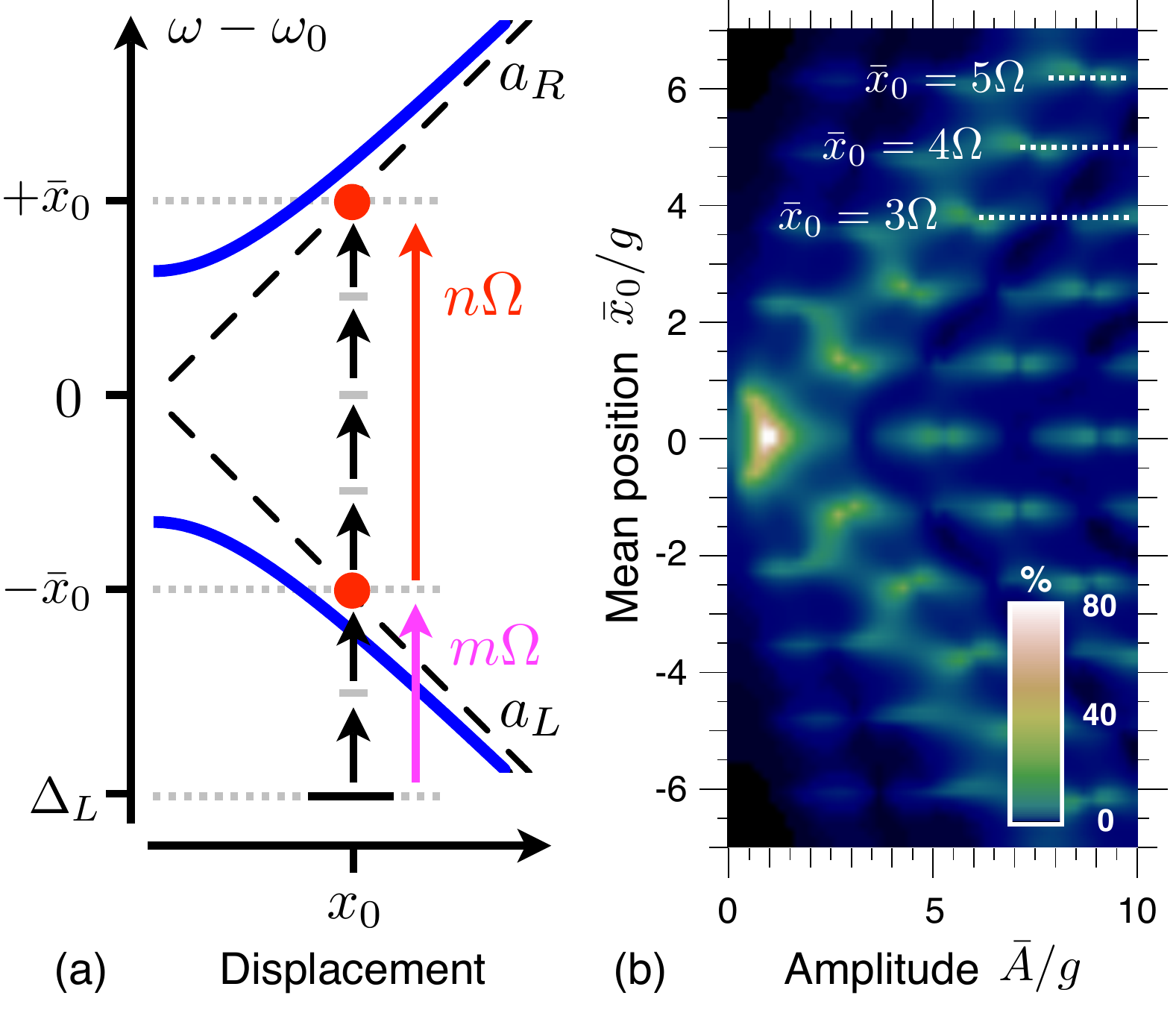}
\par\end{centering}

\caption{(a) Multiphonon transition picture. To see transmission, two processes
are involved: Firstly (magenta), excitation of the left cavity mode
by the laser drive at $\Delta_{L}$, supported by $m$ phonons. Secondly
(red), a suitable $n$-quanta multiphonon transition to transfer a
photon from the left into the right mode. (b) Density plot for time-averaged
transmission as a function of $\bar{x}_{0}$ and $\bar{A}\simeq g$.
Further parameters as in Fig.~\ref{fig:Large_amplitudes}. \label{fig:Multiphonon_SmallA}}

\end{figure}

Fig.~\ref{fig:Large_amplitudes} shows numerical results for $\bar{A}\gg\Omega,\, g$.
For experimentally accessible parameters $g/2\pi=1\,\text{MHz}$,
$l=1\,\text{cm}$ and $\omega_{0}/2\pi=3\cdot10^{14}\,\text{Hz}$,
we have $\omega_{0}/l=30g/\text{nm}$ and $\bar{A}=60g$ corresponds
to an oscillation amplitude of $A=2\,\text{nm}$ that is below the
nonlinear regime for a $50\,\text{nm}$ thick membrane. Apart from
the modulation of transmission as a function of $\bar{A}$ (see below),
we observe finite transmission only if $\bar{x}_{0}$ is a multiple
of $\Omega$. We first present an intuitive description. Transmission
is determined by two subsequent processes. First, the laser has to
excite the left mode. Secondly, the internal dynamics must be able
to transfer photons into the right one. In general, both processes
are inelastic and therefore require energy to be transferred between
the light field and the oscillating membrane. The left mode's frequency
is oscillating around the time-averaged value $-\bar{x}_{0}.$ Hence,
the resonance condition to excite the left mode reads\begin{equation}
\Delta_{L}+m\Omega=-\bar{x}_{0},\label{eq:Resonance_cond1}\end{equation}
see Fig.~\ref{fig:Multiphonon_SmallA}a. Here, $m\Omega$ is an adequate
multiphonon transition. The width of the individual resonances is
determined by $\kappa$. The subsequent process displays the physics
of LZS dynamics: LZ transitions split the photon state into a coherent
superposition, the two amplitudes gather different phases and interfere
the next time the system transverses the avoided crossing. The condition
for constructive interference can also be phrased in terms of an additional
multiphonon transition that transfers a photon from the left mode
with average frequency $-\bar{x}_{0}$ to the right one at $+\bar{x}_{0}$,\begin{equation}
n\Omega=2\bar{x}_{0}.\label{eq:Resonance_cond2}\end{equation}
We find transmission only if both conditions are met. We note that
the coupling $g$ between modes does not enter here. We will come
back to this point later.%

To derive these resonance conditions as well as to understand the
dependence on $\bar{A}$, in the following, we calculate an approximate,
analytic expression for the transmission. From Eq.~(\ref{eq:EOM_IO_theory}),
the Green's function $G(t,t')$, required for the transmission (\ref{eq:T_Greens_fct}),
is found to be \begin{equation}
G(t,t')=\tilde{a}_{R}(t,t')e^{-i\phi(t')},\label{eq:Solution_Greens}\end{equation}
where we have split off a phase $\phi(t')=(\bar{A}/\Omega)\sin(\Omega t')$,
and $\tilde{a}_{R}(t,t')$ is a solution to\begin{equation}
i\frac{d}{dt}\left(\begin{array}{c}
\tilde{a}_{R}\\
\tilde{a}_{L}\end{array}\right)=\left(\begin{array}{cc}
\bar{x}_{0} & ge^{+2i\phi(t)}\\
ge^{-2i\phi(t)} & -\bar{x}_{0}\end{array}\right)\left(\begin{array}{c}
\tilde{a}_{R}\\
\tilde{a}_{L}\end{array}\right),\label{eq:EOM_TLS}\end{equation}
with $t\geq t'$ and initial condition $\tilde{a}_{R}(t',t')=0$,
$\tilde{a}_{L}(t',t')=1$. We now show that the two multiphonon processes
introduced above correspond to the two factors in Eq.~(\ref{eq:Solution_Greens}).
The term $e^{-i\phi(t')}=\sum_{m}J_{m}(\bar{A}/\Omega)e^{-im\Omega t'}$
describes the initial excitation, where the amplitude for a transfer
of $m$ phonons is set by the Bessel function $J_{m}(\bar{A}/\Omega)$.
Secondly, the internal dynamics described by $\tilde{a}_{R}(t,t')$
is expressed in terms of a two-level system with time-dependent coupling
$ge^{2i\phi(t)}=g\sum_{n}J_{n}(2\bar{A}/\Omega)e^{in\Omega t}$. Thus,
the strength of the second multiphonon transition $n\Omega$ in Fig.~\ref{fig:Multiphonon_SmallA}a
is determined by a Bessel function $J_{n}(2\bar{A}/\Omega)$, corresponding
to Stueckelberg interferences known from atomic physics. As a special
case, this also describes the Autler-Townes splitting at small drive.
This can be calculated from (\ref{eq:EOM_TLS}) using an interaction
picture representation and considering the time-dependent coupling
only up to $J_{1}$, yielding an effective transition frequency $2\sqrt{g_{0}^{2}+\bar{x}_{0}^{2}}$,
with $g_{0}=gJ_{0}(2\bar{A}/\Omega)$, and a Rabi frequency $g_{1}=gJ_{1}(2\bar{A}/\Omega)$. 

In the case of LZS dynamics, i.e. strong drive, for sufficiently large
amplitudes only one of the harmonics of $g\sum_{n}J_{n}(2\bar{A}/\Omega)e^{in\Omega t}$
will be in resonance with the system. This corresponds to leading-order
perturbation theory within the Floquet approach \cite{Grifoni1998Driven-quantum-}
applied to Eq.~(\ref{eq:EOM_TLS}). In this case Eq.~(\ref{eq:EOM_TLS})
simplifies to the problem of a two-state system with harmonic drive
at $n\Omega$ and effective coupling constant \begin{equation}
g_{n}=gJ_{n}(2\bar{A}/\Omega).\label{eq:Effective_coupling}\end{equation}
To estimate when this approximation becomes appropriate, we note that
for a driven undamped two-state system the width of the power-broadened
resonance is set by the Rabi frequency. Thus, Eq.~(\ref{eq:EOM_TLS})
yields a series of resonance peaks at $\bar{x}_{0}=n\Omega/2$, and
they become separated if $4g_{n}<\Omega$. Using the asymptotic form
for large arguments $\bar{A}/\Omega\gg1$, $J_{n}(y)\simeq\sqrt{\frac{2}{\pi y}}\cos\left(y-\frac{n\pi}{2}-\frac{\pi}{4}\right)$,
we find the resonance approximation to hold whenever\begin{equation}
g^{2}<\frac{\pi}{16}\bar{A}\Omega.\label{eq:Condition_ResoApprox}\end{equation}
Note the resemblance to the criterion for non-adiabatic transitions
that can be derived from the standard LZ formula $P_{1\rightarrow1}=\exp(-\pi g^{2}/2v)$,
where $v=\bar{A}\Omega$ is the sweep velocity. Eq.~(\ref{eq:Condition_ResoApprox})
is clearly fulfilled for the parameters of Fig. \ref{fig:Large_amplitudes}. 

Given the resonance approximation, we find for the Green's function\begin{equation}
G(t,t')=-i\frac{g_{n}}{\omega_{n}}\sin\left(\omega_{n}(t-t')\right)e^{-in\Omega(t+t')/2}e^{-i\phi(t')},\label{eq:Solution_Greens_RA}\end{equation}
with $\omega_{n}=\sqrt{\left(g_{n}\right)^{2}+\left(\bar{x}_{0}-n\Omega/2\right)^{2}}$.
Note that $\omega_{n}$ contains $g_{n}$, which is much smaller than
the bare splitting $g$ for $\bar{A}\gg\Omega$. This explains why
the resonance conditions (\ref{eq:Resonance_cond1}) and (\ref{eq:Resonance_cond2})
involve the bare optical mode frequencies $\pm\bar{x}_{0}$ instead
of the adiabatic eigenfrequencies $\omega_{\pm}$. 

We insert (\ref{eq:Solution_Greens_RA}) into (\ref{eq:T_Greens_fct}),
taking into account the sum over independent contributions with $n$
quanta. In the resolved sideband regime ($\Omega>\kappa$), the integration
of (\ref{eq:T_Greens_fct}) selects a specific $m$ for the excitation
process, see Eq. (\ref{eq:Resonance_cond1}). Thus we find an approximate
expression for the transmission (displayed here for the special case
$\Delta_{L}=0$, where $m=2n$):\begin{eqnarray}
T & = & \left(\frac{\kappa}{g}\right)^{2}\sum_{n}\left(J_{n}\left(\frac{\bar{A}}{\Omega}\right)\times\right.\nonumber \\
 &  & \left.\frac{J_{2n}\left(2\frac{\bar{A}}{\Omega}\right)}{\frac{1}{g^{2}}\left[\left(\frac{\kappa}{2}\right)^{2}+\left(x_{0}-n\Omega\right)^{2}\right]+\left[J_{2n}\left(2\frac{\bar{A}}{\Omega}\right)\right]^{2}}\right)^{2},\label{eq:T_analytic}\end{eqnarray}
This captures fully the numerical results shown in Fig.~\ref{fig:Large_amplitudes}.
Whenever the resonance conditions are fulfilled (by choosing the offset
$\bar{x}_{0}$) , the transmission is modulated by the two Bessel
functions. While $J_{m}(\bar{A}/\Omega)$ describing the excitation
process depends on the amplitude $\bar{A}$, the LZS dynamics characterized
by $J_{n}(2\bar{A}/\Omega)$ is determined by the phase difference
gathered between LZ transitions, involving $2\bar{A}$. According
to Eq. (\ref{eq:Resonance_cond1}), if we were to increase $\Delta_{L}$
in Fig.~\ref{fig:Large_amplitudes} we would tune out of resonance
and the transmission would vanish everywhere. For $\Delta_{L}=\Omega/2$,
the conditions (\ref{eq:Resonance_cond1}) and (\ref{eq:Resonance_cond2})
can be met for $\bar{x}_{0}$ being an \emph{odd} multiple of $\Omega/2$.
For $\Delta_{L}=\Omega$ we would again find transmission for $\bar{x}_{0}$
being a multiple of $\Omega$. Note however that the entire plot would
be shifted in $\bar{A}$ by an amount $\pi\Omega/2$ due to the changed
index of the Bessel function $J_{m}$. Finally, at $\Delta_{L}=2\Omega$
we would recover Fig. \ref{fig:Large_amplitudes}.

Fig.~\ref{fig:Multiphonon_SmallA}b shows numerical results for smaller
values of $\bar{A}/g$ (while keeping $\Omega$ as in Fig.~\ref{fig:Large_amplitudes}).
As before, we see resonances for $\bar{x}_{0}$ being a multiple of
$\Omega$ and expect regions of excitation with width $\kappa$ determined
by $J_{m}(\bar{A}/\Omega)$. Within these regions, we note the already
familiar substructure that is due to LZS dynamics.%

To conclude, we proposed a setup to investigate non-equilibrium photon
dynamics driven by mechanical motion in an optomechanical system with
a membrane inside a cavity. We predicted the possibility to observe
Autler-Townes splitting and features of Landau-Zener-Stueckelberg
dynamics in the transmission spectrum. The observation of the effects
discussed here is within reach of current experiments. The same nontrivial
light field dynamics will enter when describing self-induced nonlinear
optomechanical oscillations in these systems, which would be an interesting
topic for future research.

We acknowledge fruitful discussions with Jack Sankey, as well as support
by the DFG (SFB 631, SFB TR 12, NIM, Emmy-Noether program), DIP, and
(J.H.) via NSF 0855455, NSF 0653377, and AFOSR FA9550-09-1-0484.

\bibliographystyle{apsrev}
\bibliography{Bib_LZ_25cites}

\end{document}